\documentclass[aps,pra,twocolumn,showpacs]{revtex4-1}

\usepackage{graphicx,epstopdf}
\usepackage{amsmath}
\usepackage{amssymb}
\usepackage{braket}
\usepackage{color}
\usepackage{float}
\usepackage{epstopdf}
\usepackage{amsmath}
\usepackage{braket}
\usepackage{bm}
\usepackage{graphics,epstopdf}
\usepackage{color}
\usepackage[usenames,dvipsnames,svgnames]{pstricks}
\usepackage{epsfig}
\usepackage{pst-grad} 
\usepackage{pst-plot}

\usepackage[colorlinks=true, citecolor=blue, urlcolor=blue ]{hyperref}
\input{epsf}
\begin{document}

%
%

\title{Polarization squeezing in polarized light}
\author{Ranjana Prakash and Namrata Shukla} 
\affiliation{Physics Department, University of Allahabad, Allahabad, UP 211002, India} 

\date{\today}

\begin{abstract}
 It is shown that polarized light can be polarization squeezed only if it exhibits sub-Poissonian statistics with the Mandel's $Q$ factor less than -1/2.
\end{abstract}

\maketitle
In classical optics, Stokes parameters are used to denote the polarization state \cite{stokes,bornwolf}. For light beam travelling along the 3-direction, the Stokes parameters $S_{0,1,2,3}$ are defined by
\begin{equation}
\label{eq1}
S_{0,1}= \langle{\mathcal{E}}_{x}^{*}{\mathcal{E}}_{x}\rangle
\pm\langle{\mathcal{E}}_{y}^{*}{\mathcal{E}}_{y}\rangle,~
S_{2}+i S_{3}=2 \langle{\mathcal{E}}_{x}^{*}{\mathcal{E}}_{y}\rangle,
\end{equation}
where $ {\mathcal E}=\bm e_x \mathcal{E}_x+\bm e_y \mathcal{E}_y$ is the analytic signal \cite{bornwolf1} for electric field and conical brackets denote the average. For perfectly polarized light, 
$S_{0}^2=|\bm S|^2=S_{1}^2+S_{2}^2+S_{3}^2$, and the point with coordinates $(S_1, S_2, S_3)$ lies on a sphere of radius $S_0$, called the Poincare's sphere \cite{bornwolf2}, and the direction of $\bm S = (S1, S2, S3)$ represents the polarization state. For unpolarized light \cite{bornwolf3, prakash, prakash1, prakashagarwal}, $\bm S = 0$ and for partially polarized light $|\bm S| < S_0$ and the point with coordinates $(S1, S2, S3)$ lies inside the Poincare sphere. Since the Stokes parameters involve only the coherence functions \cite{mandelwolf} of order $(1, 1)$, they are not sufficient for describing polarization in the context of nonlinear interactions (also discussed in ref. \cite{prakash}). Quantum analogue of Stokes parameters are the Stokes operators defined by
\begin{equation}
\label{eq2}
\hat S_{0, 1}=\hat a_{x}^\dagger \hat a_{x}\pm \hat a_{y}^\dagger \hat a_{y},~ \hat S_{2}+i \hat S_{3}=2 \hat a_{x}^\dagger \hat a_{y}.
\end{equation}
where $\hat a_{x, y}$ are the annihilation operators for the $x$ and $y$ linear
polarization. Stokes operators satisfy the commutation relations,

\begin{equation}
\label{eq3}
[\hat S_0, \hat S_j]=0,~[\hat S_j, \hat S_k]=2i {\sum_{l}\epsilon_{jkl}}~\hat S_{l},~ (j,k,l=1,2,3),
\end{equation}
and lead to the uncertainty relations,
\begin{eqnarray}
\label{eq4}
V_{j} V_{k}\geqslant{\langle\hat S_{l}\rangle}^2,~ 
V_{j}\equiv\langle\hat S_{j}^2\rangle- {\langle\hat S_{j}\rangle}^2,~(j\neq k\neq l\neq j).
\end{eqnarray}

Relations for Stokes operators are very much similar to those for Dicke's collective atom hermitian operators \cite{dicke} $\hat R_{1,2,3}$ for two-level atoms (TLA's). If $|u\rangle_j$and $|l\rangle_j$ are upper and lower states for the $j^{th}$ TLA, for an assembly of $N$ TLA's the Dicke's operators $\hat R_{1,2,3}$ are defined by
\begin {equation}
\label{eq5}
\hat R_1+i \hat R_2=\sum_{j=1}^{N} |u\rangle_j \langle l|,~ \hat R_3=  \sum_{j=1}^{N} \frac{1}{2}[||u\rangle_j \langle u|-|l\rangle_j \langle l|],
\end {equation}
and satisfy
\begin {equation}
\label{eq6}
[\hat R_1,\hat R_2]=i{\sum_{l}\epsilon_{jkl}}~\hat R_{l},~ (j,k,l=1,2,3),
\end {equation}
which is similar to those in Eq.\eqref{eq3}, except for the factor of 2 on right hand side. These lead to uncertainty relations on the basis of which Walls and Zoller \cite{walls} defined atomic squeezing of $\hat {R_1}$ if
\begin{equation}
\label{eq7}
{\langle\hat {R_1}}^2\rangle-\langle\hat {R_1}\rangle^2<\frac{1}{2}|\langle\hat {R_2}\rangle|~ or~ \frac{1}{2}|\langle\hat {R_3}\rangle|.
\end{equation}
This was generalized by Prakash and Kumar \cite{prakash2}, who call the  generalized component $\hat R_{\bm n}= ({\bm n}.\hat {\bm R})$ of $\hat {\bm R}=(\hat R_1, \hat R_2, \hat R_3)$ along the unit vector ${\bm n}$ squeezed if
\begin{equation}
\label{eq8}
\langle\hat R_{\bm n}^2\rangle-\langle\hat R_{\bm n}\rangle^2\leq\frac{1}{2}{\bigg[{\langle\hat R_{\bm n_{\perp 1}}\rangle}^2 +{\langle\hat R_{\bm n_{\perp 2}}\rangle}^2\bigg]}^{1/2}.
\end{equation}
where $\bm n_{\perp 1}$ and $\bm n_{\perp 2}$ are any two unit vectors perpendicular to $\bm n$.

For optical polarization, concept of polarization squeezing is introduced through 
 Eq.\eqref{eq3}. Chirkin et al. \cite{chirkin} gave the first definition in the form, 
$ V_j<V_j(coh)=\langle \hat S_0\rangle$, for $j=1,2$ or 3 where $ V_j=\langle \hat S_j^2\rangle-\langle \hat S_j\rangle^2 $ is variance of the operator $\hat S_j$ and $\hat V_j(coh)$ is the variance of equally intense coherent state. Heersink et al. \cite{heersink} used Eq.\eqref{eq3} and called operator $\hat S_j$ polarization squeezed if $V_j<|\langle\hat S_l\rangle|<V_k $, for $j\neq k\neq l\neq j$, which is similar to the definition in Eq.\eqref{eq7} of Walls and Zoller \cite{walls} for atomic squeezing. This was generalized by Luis and Korolkova \cite{luis} who wrote criterion for squeezing as

\begin{equation}
\label{eq9}
V_{\bm n}<|\langle\hat S_{\bm n \perp}\rangle|,~ V_{\bm n}\equiv\langle\hat S_{\bm n}^2\rangle-\langle\hat S_{\bm n}\rangle^2,
\end{equation}
where $\bm n_\perp$ is a unit vector perpendicular to $\bm n$. This was written by Prakash and Shukla \cite {prakashshukla} in the form
\begin{eqnarray}
\label{eq10}
V_{\bm n}\equiv\langle \hat S_{\bm n}^2\rangle - {\langle \hat S_{\bm n}\rangle}^2<{|\langle \hat S_{\bm n \perp\rangle} |}_{max}\nonumber\\
={\bigg[{|\langle \hat {\bm S} \rangle |}^2-{\langle \hat S_{\bm n} \rangle}^2 \bigg]}^{1/2},
\end{eqnarray}
which is very much similar to definition in Eq.\eqref{eq8} for atomic squeezing by prakash and kumar \cite{prakash2}. 

The basis states for study of polarization need not necessarily be the two linear polarizations and can in the most general case be the two general orthogonal elliptical polarizations represented by two orthogonal unit vectors, say,  $\bm\epsilon$ and ${\bm\epsilon}_\perp$which satisfy obviously $\bm\epsilon^{*}.\bm\epsilon=\bm\epsilon_\perp^{*}.\bm\epsilon_\perp=1$ and $\bm\epsilon^{*}.\bm\epsilon_\perp=0$. Since we can write the expansion of vector potential $\mathcal {\hat A}$ for a monochromatic unidirectional optical field in the form,
\begin{eqnarray}
\label{eq11}
\mathcal{\hat A}=\sqrt{\frac{2\pi}{{\omega \mathcal V}}}\bigg[\big(\bm \epsilon {\hat a}_{\bm \epsilon} 
+\bm \epsilon _\perp \hat a_{\bm \epsilon_{\perp}}\big) e^{i kz}+h.c.\bigg] \nonumber\\
=\sqrt{\frac{2\pi}{{\omega \mathcal V}}}\bigg[\big(\bm e_x {\hat a}_{x} 
+\bm e_y \hat a_{y}\big) e^{i kz}+h.c.\bigg]
\end{eqnarray}
in natural units, where $h.c.$ stands for hermitian conjugate, it leads to [see, {\it e.g}, Ref. \cite {prakash1} also]

\begin{equation}
\label{eq12}
\hat a_{\bm \epsilon}=\bm \epsilon_x^{*} \hat a_x+\bm \epsilon_y^{*} \hat a_y,~ \hat a_{\bm \epsilon_{\perp}}=\bm \epsilon_{\perp x}^{*} \hat a_x+\bm \epsilon_{\perp y}^{*} \hat a_y,
\end{equation}
and also
\begin{equation}
\label{eq13}
\hat a_x={\bm \epsilon}_x \hat a_{\bm \epsilon}+{\bm \epsilon}_{\perp x} \hat a_{{\bm\epsilon}_{\perp}},~ \hat a_y={\bm \epsilon}_y \hat a_{\bm \epsilon}+{\bm \epsilon}_{\perp y}\hat a_{{\bm\epsilon}_{\perp}}.
\end{equation}

Operators $\hat a_{\bm \epsilon}$ and $ \hat a_{{\bm \epsilon}_{\perp}}$ are annihilation operator for two orthogonal modes having polarization represented by complex unit vector $\bm \epsilon$ and $\bm \epsilon_\perp$. These help us define \cite{prakashsingh}  state $|\psi\rangle $ of light polarized in the mode $\bm \epsilon$ by

\begin{equation}
\label{eq14}
\hat a_{{\bm \epsilon}_{\perp}} |\psi\rangle=0,~ \langle\psi| \hat a_{{\bm \epsilon}_{\perp}}^\dagger=0.
\end{equation}

To study polarization squeezing in this state, straight calculations using Eq.\eqref{eq2} and Eq.\eqref{eq14}give
\begin{eqnarray}
&&\langle \hat S_0\rangle=\langle \hat a_{\bm \epsilon}^\dagger \hat a_{\bm \epsilon}\rangle,~\langle \hat S_j\rangle=m_j\langle \hat a_{\bm \epsilon}^\dagger \hat a_{\bm \epsilon} \rangle\label{eq15}\\
&&\langle \hat S_j^2\rangle=\langle \hat a_{\bm \epsilon}^\dagger \hat a_{\bm \epsilon}\rangle+m_j^2\langle\hat a_{\bm \epsilon}^{\dagger 2} \hat a_{\bm \epsilon}^{2}\rangle\label{eq16}\\
&&\big\langle\big\{\hat S_j, \hat S_k\big\}\big\rangle=2m_{j}m_k\langle\hat a_{\bm\epsilon}^{\dagger 2} \hat a_{\bm \epsilon}^2\rangle, ~~~(j\neq k),\label{eq17}
\end{eqnarray}
where 
\begin{eqnarray}
\label{eq18}
&&m_1={|{\bm \epsilon_x}|}^2-{|{\bm \epsilon_y}|}^2,\nonumber\\
&&m_2={\bm \epsilon}_x^{*} {\bm \epsilon}_y+{\bm \epsilon}_y^{*}{\bm \epsilon}_x,~m_3=(-i {\bm \epsilon}_x^{*}{\bm \epsilon}_y+i {\bm \epsilon}_y^{*}{\bm \epsilon}_x),
\end{eqnarray} 
define a unit vector $\bm m$. If we write
\begin{equation}
\label{eq19}
\bm \epsilon= {\bm e}_x \cos \frac{\theta_0}{2}+{\bm e}_y \sin \frac{\theta_0}{2} e^{i \Phi_0},
\end{equation}
and let angles $\theta_0$ and $\Phi_0$ define the polarization state, the polarization state will also be represented by unit vector,
\begin{equation}
\label{eq20}
\bm m={\bm e}_x\cos\theta_0+({\bm e}_y \cos \Phi_0+{\bm e}_z\sin \Phi_0)\sin\theta_0.
\end{equation}

We can write the unit vector $\bm n$, squeezing of components of $S$ along which we are considering, in a similar form as
 \begin{equation}
\label{eq21}
\bm n={\bm e}_x\cos\theta+({\bm e}_y \cos \Phi+{\bm e}_z\sin \Phi)\sin\theta
\end{equation}
Eqs. \eqref{eq15}-\eqref{eq17}, then give
\begin{equation}
\label{eq 22}
\langle \hat S_{\bm n}\rangle=\langle \hat a_{\bm \epsilon}^\dagger \hat a_{\bm \epsilon}\rangle \cos\Phi,~\langle \hat S_{\bm n}^2\rangle=\langle \hat a_{\bm \epsilon}^\dagger \hat a_{\bm \epsilon}\rangle +\langle\hat a_{\bm \epsilon}^{\dagger 2} \hat a_{\bm \epsilon}^{2}\rangle\cos^{2}\Phi,
\end{equation}
where
\begin{equation}
\label{23}
\cos \Phi=(\bm n.\bm m)=\cos\theta_0\cos\theta+\cos(\Phi_0-\Phi)\sin\theta_0\sin\theta.
\end{equation}
Here $\Phi$ is angle between the unit vectors $\bm n$ and $\bm m $ with $0\leq\Phi\leq\pi$. Polarization squeezing therefore occurs if
\begin{eqnarray}
\label{eq24}
&&\langle \hat S_{\bm n}^2\rangle - {\langle \hat S_{\bm n}\rangle}^2-{\bigg[{|\langle \hat {\bm S} \rangle |}^2-{\langle \hat S_{\bm n} \rangle}^2 \bigg]}^{1/2} \nonumber\\
&&=\langle \hat a_{\bm \epsilon}^\dagger \hat a_{\bm \epsilon}\rangle (1-\sin\Phi)+ \big[\langle\hat a_{\bm \epsilon}^{\dagger 2} \hat a_{\bm \epsilon}^{2}\rangle-\langle \hat a_{\bm \epsilon}^\dagger \hat a_{\bm \epsilon}\rangle^2 \big]\cos^{2}\Phi \nonumber\\
&&<0.
\end{eqnarray}
Mandel's $Q$ factor is defined by \cite{mandelwolf1}
\begin{equation}
\label{eq25}
Q=\big(\langle\hat a_{\bm \epsilon}^{\dagger 2} \hat a_{\bm \epsilon}^{2}\rangle-\langle \hat a_{\bm \epsilon}^\dagger \hat a_{\bm \epsilon}\rangle^2 \big).
\end{equation}
For classical fields, $Q\geq0$. $Q\leq0$ gives the non-classical features of light, sub-Poissonian statistics. The criterion for polarization squeezing is then,
\begin{equation}
\label{eq26}
1-\sin\Phi+Q\cos^2\Phi=(1-\sin\Phi)(1+Q[1+\sin\Phi])<0
\end{equation}
This cannot be satisfied for $Q\geq0$, {\it i.e.} for Poissonian or super-Poissonian statistics. For $Q<0$ also, this cannot be satisfied if $Q\geq -1/2$. For $Q<-1/2$, however, this can be satisfied for values of $\Phi$ for which $1>\sin\Phi >|Q|^{-1}-1$, which is same as $ 0<\cos\Phi<{[2|Q|-1]}^{1/2}/|Q|$. Polarized light in the state represented by $\bm m$ can thus be polarization squeezed in Stokes operator $\hat S_{\bm n}$ only if it exhibits sub-Poissonian statistics and $Q<-{(1+\sin\Phi)}^{-1}$. For a given value of $Q$ which is less than $-1/2$, we thus get a cone of semi-vertical angle $\sin^{-1}({|Q|}^{-1}-1)$ about the unit vector $\bm m$ which describes the polarization state. If the line of $\bm n$ lies outside this cone and is not perpendicular to its axis then $\hat S_{\bm n}$ is squeezed.

It is also interesting to see that if $Q=-1$, the lowest possible value, which occurs for pure photon number state, the semi-vertical angle of cone is zero and hence all components $\hat S_{\bm n}$ are squeezed except those for which $\Phi=0$ or $\pi/2$ \cite{prakashshukla}.

If we use the angles $\theta$ and $\Phi$, {\it i.e.}, unit vector $\bm n$, to define orthogonal complex unit vectors $\bar{{\bm \epsilon}}$ and $\bar{{\bm \epsilon}}_\perp$ by 
\begin{eqnarray}
\label{27}
&&\bar{{\bm \epsilon}}=\cos\frac{\theta}{2}{\bm e}_x+\sin\frac{\theta}{2}e^{i \Phi}{\bm e}_y, \nonumber\\
&&\bar{{\bm \epsilon}}_\perp=-\sin\frac{\theta}{2}{\bm e}_x+\cos\frac{\theta}{2}e^{i \Phi}{\bm e}_y,
\end{eqnarray}
it can be shown that 
\begin{equation}
\label{eq28}
\hat S_{\bm n}=(\bm n. \hat{\bm S})=\hat N_{{\bm \epsilon}}-\hat N_{{\bm \epsilon}_ {\perp}},~ 
{\langle{\hat S}\rangle}^2-{\langle{\hat S}_{\bm n}\rangle}^2=4 \langle \hat N_{\bar{{\bm \epsilon}}}\rangle \langle\hat N_{\bar{{\bm \epsilon}}_{\perp}}\rangle,
\end{equation}
where $\hat N_{\bar{{\bm \epsilon}}}=\hat a_{\bar{{{\bm \epsilon}}}}^{\dagger} \hat a_{\bar{{\bm \epsilon}}}$ and $\hat N_{\bar{{\bm \epsilon}}_{\perp}}=\hat a_{\bar{{\bm \epsilon}}_{\perp}}^{\dagger}\hat a_{\bar{{\bm \epsilon}}_{\perp}}$ are photon number operators for light polarized along $\bar{{\bm \epsilon}}$ and $\bar{{\bm \epsilon}}_{\perp}$. Eq.\eqref{eq28}, helps us write
\begin{eqnarray}
\label{eq29}
&&V_{\bm n}-{\bigg[{|\langle \hat {\bm S} \rangle |}^2-{\langle \hat S_{\bm n} \rangle}^2 \bigg]}^{1/2}\nonumber\\
&&=\langle\hat N_{\bar{{\bm \epsilon}}}^2\rangle+\langle\hat N_{\bar{{\bm \epsilon}}_{ \perp}}^2\rangle-2\langle\hat N_{\bar{{\bm \epsilon}}}\hat N_{\bar{{\bm \epsilon}}_{\perp}}\rangle
-{( \langle \hat N_{\bar{{\bm \epsilon}}}\rangle-\langle \hat N_{\bar{{\bm \epsilon}}_{\perp}}\rangle\big)}^2\nonumber\\
~~~~~~~&&-2\langle \hat N_{\bar{{\bm \epsilon}}}\rangle^{1/2} \langle\hat N_{\bar{{\bm \epsilon}}_ {\perp}}\rangle^{1/2}
\end{eqnarray}
This shows that, to detect squeezing in $\hat S_{\bm n}$, therefore, only measurement of expectation values of $\hat N_{\bar{{\bm \epsilon}}}$ and $\hat N_{\bar{{\bm \epsilon}} _{\perp}}$ and their squares and product is required. This can be done easily by shifting phase of y-component by $\Phi$ followed by rotating the beam by $-\theta/2$ about the direction of propagation and measurement in $x$ and $y$ linearly polarized components. 
Also since Eq.\eqref{eq29} can be written as 
\begin{eqnarray}
&&V_{\bm n}-{\bigg[{|\langle \hat {\bm S} \rangle |}^2-{\langle \hat S_{\bm n} \rangle}^2 \bigg]}^{1/2}\nonumber\\
&&=\langle\hat a_{\bar{{\bm \epsilon}}}^{\dagger 2}\hat a_{\bar{{\bm \epsilon}}}^{\dagger 2}\rangle +\langle\hat a_{\bar{{\bm \epsilon}}_{\perp}}^{\dagger 2}\hat a_{\bar{{\bm \epsilon}}_ {\perp}}^{\dagger 2}\rangle - 2\langle\hat a_{\bar{{\bm \epsilon}}}^{\dagger}\hat a_{\bar{{\bm \epsilon}}_{\perp}}^{\dagger} \hat a_{\bar{{\bm \epsilon}}}\hat a_{\bar{{\bm \epsilon}}_ {\perp}}\rangle \nonumber\\
&&-{\big[\langle \hat a_{\bar{{\bm \epsilon}}}^\dagger \hat a_{\bar{{\bm \epsilon}}}\rangle -\langle \hat a_{\bar{{\bm \epsilon}}_{\perp}}^{\dagger 2} \hat a_{\bar{{\bm \epsilon}}_{\perp}}^2\rangle \big]}^2+{\big[{\langle \hat a_{\bar{{\bm \epsilon}}}^\dagger \hat a_{\bar{{\bm \epsilon}}}\rangle}^{1/2}-{\langle \hat a_{\bar{{\bm \epsilon}}_{\perp}}^{\dagger 2} \hat a_{\bar{{\bm \epsilon}}_{\perp}}^{2}\rangle}^{1/2}\big]}^2,\nonumber\\
\label{30}
\end{eqnarray}
If the density operator of light is written in the Sudarshan-Glauber diagonal representation \cite{sudarshan} in the basis of coherent state $|\alpha, \beta\rangle_{{\bm\epsilon}, {\bm \epsilon}_\perp}$ in the form

\begin{equation}
\label{eq31}
\rho=\int{ d^2\alpha~d^2\beta~P(\alpha, \beta) \ket{\alpha, \beta}_{{\bm \epsilon},{\bm \epsilon}_{\perp}} \bra{\alpha, \beta}},
\end{equation}
we have
\begin{eqnarray}
\label{eq32}
&&V_{\bm n}-{\bigg[{|\langle \hat {\bm S} \rangle |}^2-{\langle \hat S_{\bm n} \rangle}^2 \bigg]}^{1/2}\nonumber\\
&&=\int d^2\alpha~d^2\beta~P(\alpha, \beta)\bigg[\big \{|\alpha|^2-|\beta|^2\nonumber\\
&&~~~~-(\langle|\alpha |^2\rangle-\langle |\beta |^2\rangle)\big \}^2+{(|\alpha |-|\beta |)}^2\bigg].
\end{eqnarray}
This shows that if polarization squeezing is exhibited, no non-negative $P(\alpha, \beta)$ can exist and therefore this is a purely non-classical feature.
\vspace{0.2cm}\section*{Acknowledgement}
We would like to thank Hari Prakash and Naresh Chandra for their interest and critical comments.


\begin{thebibliography}{0}
\bibitem{stokes}
G. G. Stokes, {\it Trans. Cambridge. Philos. Soc.}{\bf 9} 399 (1852) .
\bibitem{bornwolf}
M. Born, E. Wolf, {\it Principles of Optics} (Cambridge University Press, England, 1999) p. 30.
\bibitem{bornwolf1}
 See, e.g., ref. [2], pp. 494-499.
\bibitem{bornwolf2}
 See, e.g., ref. [2], p. 31.
\bibitem{bornwolf3}
See, e.g., ref. [2], p. 423.
\bibitem{prakash}
H. Prakash, N. Chandra, {\it Phys. Rev. A} {\bf 4} 796 (1971).
\bibitem{prakash1}
 H. Prakash, N. Chandra, {\it Phys. Lett. A} {\bf 34} 28 (1971).
\bibitem{prakashagarwal} 
H. Prakash, N. Chandra, {\it Phys. Rev. A} {\bf 9} 1021(1974) ; 
H. Prakash, N. Chandra, {\it Phys. Lett. A} {\bf 31}  331 (1970); 
H. Prakash, D. K. Singh, {\it J. Phys. B} {\bf 41} 4 (2008); 
G. S. Agarwal, {\it Lett. Nuovo Cimento} {\bf 53} 97 (1970).
\bibitem{mandelwolf} 
L. Mandel, E. Wolf, {\it Optical Coherence and Quantum Optics} (Cambridge University Press, England, 1995) p. 423.
\bibitem{dicke} 
R. H. Dicke, {\it Phys. Rev.} {\bf 93} 99 (1954).
\bibitem{walls}
 D. F. Walls, P. Zoller, {\it Phys. Rev. Lett.} {\bf 47} 709 (1981).
\bibitem{prakash2}
H. Prakash, R. Kumar, {\it J. Opt. B: Quantum and Semiclass. Opt.} {\bf 7} S757 (2005)
\bibitem{chirkin}A. S. Chirkin, A. A. Orlov, D. Y. Paraschuk, {\it Kvant. Elektron} {\bf 20} 999 (1993); A. S. Chirkin, A. A. Orlov, D.Y. Paraschuk, {\it Quantum Electron} {\bf 23} 870 (1993).
\bibitem{heersink}
J. Heersink, T. Lorenz, O. Glockl, N. Korolkova, G. Leuchs, {\it Phys. Rev. A.} {\bf 68} 013815 (2003).
\bibitem{luis}
 A. Luis, N. Korolkova, {\it Phys. Rev. A} {\bf 74} 043817 (2006).
\bibitem{prakashshukla}
Ranjana Prakash and Namrata Shukla, {\it Proceedings of International Conference on Optics and Photonics (ICOP 2009)} , p. 244 (2009).
\bibitem{prakashsingh}
H. Prakash, R. S. Singh, {\it J. Phys. Soc. Jpn} {\bf 69} 284 (2000).
\bibitem{mandelwolf1}
See, e.g., Ref. [9], p. 627.
\bibitem{sudarshan} E. G. Sudarshan, {\it Phys. Rev. Lett.} {\bf 10}  277 (1963); R. J. Glauber, {\it Phys. Rev. Lett.} {\bf 130} 2529 (1963).

\end{thebibliography}
\end{document}